\documentclass[prl,aps,twocolumn,floatfix,showpacspsfig]{revtex4}
\usepackage{amssymb}

\usepackage{epsfig}

\newcommand{\be}{\begin{equation}}
\newcommand{\ee}{\end{equation}}
\newcommand{\bea}{\begin{eqnarray}}
\newcommand{\eea}{\end{eqnarray}}

\newcommand{\p}{\partial}
\newcommand{\s}{\sigma}

\newcommand{\la}{\langle}
\newcommand{\ra}{\rangle}
\newcommand{\rd}{\mbox{d}}
\newcommand{\ri}{\mbox{i}}
\newcommand{\re}{\mbox{e}}

\newcommand{\Tt}{{\tilde\theta}}
\newcommand{\tp}{{\theta}}

\begin{document}
\title{Finite temperature correlation function for one-dimensional Quantum Ising model: the virial expansion. }
\author{S. A. Reyes and A. M. Tsvelik$^\dagger$}
\affiliation{ Department of Physics and Astronomy, SUNY at Stony Brook, Stony Brook, NY 11790-3800, USA;\\ 
 $^\dagger$Department of  Condensed Matter Physics and Materials Science, Brookhaven National Laboratory, Upton, NY 11973-5000, USA}
\date{\today}

\begin{abstract}
 We rewrite the exact expression for the  finite temperature two-point correlation function for the magnetization as a partition function of some field theory. This removes singularities and provides a convenient form to develop a virial expansion (the expansion in powers of soliton density).

\end{abstract}

\maketitle
\narrowtext

\sloppy
 
 To calculate correlation functions in strongly correlated systems is not an easy task, even if the corresponding models happen to be integrable. 
For models with dynamically generated spectral gaps the most powerful technique is the formfactor approach pioneered  by Karowski {\it et. al.} \cite{karowski} and perfected by Smirnov \cite{smirnov}. This approach works wonderfully for zero temperature, but  encounters difficulties at $T \neq 0$. These difficulties are related to singularities in the operator matrix elements (formfactors). These singularities exist for operators nonlocal with respect to solitons, they originate from forward scattering processes and their treatment requires careful infrared regularization. Despite long efforts a correct regularization  has not been yet found. However, for models of free fermions (such as the XY model or the Quantum Ising model), there are alternative means to calculate the correlation functions which allow to bypass the above problems. These alternative approaches include the determinant representation of the correlation functions \cite{its},\cite{korepin} and the semiclassical method \cite{sachdev} (which may have much wider application, see \cite{rapp},\cite{modsubir}). For these results to have a greater use one has to establish their relationship with the formfactor approach. A step in this direction was made in   \cite{akt} where the semiclassical results \cite{sachdev},\cite{rapp},\cite{modsubir} were reproduced by summing up the leading singularities in the formfactor expansion.  Such summation was restricted to the leading order in the soliton density $n \sim \exp(- M/T)$ ($M$ is the spectral gap). In this paper we describe  a formfactor-based representation for correlation functions which, though in its present form is valid only for models of free fermions, is rather suggestive and may give rise to useful generalizations in the future. For the  Quantum Ising (QI) model, which is the main object of this paper, this procedure naturally gives rise to  a virial expansion of the dynamical spin susceptibility. 

  The QI model  Hamiltonian is  
\bea
 H = \sum_n[- J\s^z_n\s^z_{n+1} + h\s^x_n] \label{is}
\eea
where $s^z,s^x$ are the Pauli matrices. By the Jordan-Wigner transformation this Hamiltonian can be transformed into the Hamiltonian of non-interacting fermions:
\bea
H = \sum_p \epsilon(p)F^+_pF_p ;~~ \epsilon(p) = \sqrt{(J- h)^2 + 4Jh\sin^2(p/2)}, \label{isF}
\eea
Hamiltonian (\ref{is})  possesses a property of self-duality: the transformation $\mu^z_{n+1/2} = \prod_{j<n}\s_j^x, \mu^x_{n+1/2} = \s^z_n\s^z_{n+1}$ preserves both  the commutation relations and the form of the Hamiltonian: 
\bea
H = \sum_n[- h\mu^z_{n-1/2}\mu^z_{n+1/2} + J\mu^x_{n+1/2}]
\eea
 As follows from the form of the dispersion law (\ref{isF}), $h=J$ is the critical point. It is easy to see that at $T=0$ it separates the regions where $\la \s^z\ra \neq 0 (J >h)$ and $\la\mu^z_{n+1/2}\ra  \neq 0 (J < h)$. Operators $\s^z$ and $\mu^z$ are respectively called order and disorder parameter operators. The duality allows to study the correlation functions at one side of the transition only. For instance, a correlation function of $\s^z$ at $J > h$ coincides with the correlation function for $\mu^z$ with $J,h$ interchanged. 

 Though one cannot observe a single fermion, the fermionic statistics can be indeirectly tested by measuring the correlation functions of $\s^x$. The operator $\s^x $ (as well as $\mu^x$) is  local in fermions: 
\bea
&& \s^x(x) = \sum_k\re^{-\ri q x}\gamma(k)\gamma(k-q)\hat F_k\hat F_{q -k}, ~~ \hat F_k = 
\hat F^+_{-k}, \nonumber\\
&& \gamma(k) = \sqrt{1 + k/\epsilon(k)}.\label{edensity}
\eea
Its  finite temperature correlation functions are just polarization loops; they clearly contain the Fermi distribution functions of the fermion fields $F$. Since the order  ($\s^z$) and disorder parameter ($\mu^z$) fields are nonlocal in terms of fermions, their correlation functions are more complicated.

  The duality allows us to consider $h > J$ or $h < J$ phase only. In this paper we will study the two-point correlation functions of the $\s^z$ and $\mu^z$ operators in the ``ordered'' phase $h < J$.  For technical reasons it will be convenient to work in the limit $|J -h| << J$ when the spectral gap is much smaller than the bandwidth and one can formulate a continuous description. In this limit the excitation spectrum is relativistic $\epsilon(p) = \sqrt{c^2p^2 + M^2}$, $c^2 = Jh$. Energy and momentum of a quasi-particle are conveniently parameterized by 
a rapidity, $\theta$, ($cp = M\sinh\theta$). Then the eigenstates of Hamiltonian (\ref{is}) are 
labeled  by sets of 
rapidities, $\{\theta_i\}$, such that the energy and momentum of the system are equal to
\bea
E = M\sum_{i=1}^n\cosh\theta_i, ~~ P = c^{-1}M\sum_{i=1}^n\sinh\theta_i.
\eea
Below we set $c=1$. 
  
A convenient finite temperature expression for the  two point correlation functions of $\s$ and $\mu$ was derived by Bourgij and Lisovyy\cite{kiev}. This expression for the Matsubara time correlation function is manifestly free of singularities and has the following form:   
\bea
&&\la\s(\tau,x)\s(0,0)\ra = 
CM^{1/4}\re^{-|x|\Delta(T)}\times\label{kiev1}\\
&&\sum_{N=0}^{\infty}\frac{T^{2N}}{(2N)!}
\sum_{q_1,...q_{2N}}\prod_{i=1}^{2N}
\frac{\re^{-|x|\epsilon_i - \ri \tau q_i - \eta(q_i)}}{\epsilon_i}\prod_{i > j}
\left(\frac{q_i - q_j}{\epsilon_i + \epsilon_j}\right)^2\nonumber
\eea
where $\tau$ is imaginary time. The same  expression holds for $\mu^z$, but with $2N$ replaced by $2N +1$.  $q =2\pi Tm$ ($m$ integer), and $\epsilon(q) = \sqrt{M^2 + q^2}$. The term in the exponent is ($\beta = 1/T$) 
\bea
\eta(q) = 
\frac{2\epsilon(q)}{\pi}\int_0^{\infty}\frac{\rd x}{\epsilon^2(q) + x^2}\ln\coth[\beta\epsilon(x)/2],
\eea
and 
 \bea
\Delta(T) = \int_{-\infty}^{\infty}\frac{\rd p}{\pi}\ln\{\coth[\beta\epsilon(p)/2]\} 
\eea
The symmetry breaking transition at $T=0$ leads to a finite magnetization,  
$\la\s\ra = \pm [CM^{1/4}]^{1/2}$. This is reflected in the zeroth order term in Eq. (\ref{kiev1}).

\begin{figure}
[ht]
\begin{center}
\epsfxsize=0.2\textwidth
\epsfbox{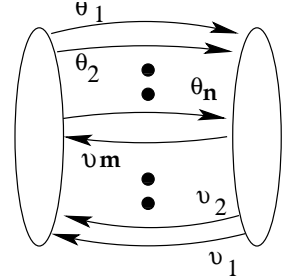}
\end{center}
\caption{A graphic representation of Eq.(\ref{truth}). The ellipsises are formfactors of $\s$ operator. Lines with left arrows are $f^{(+)}(\theta)\exp[\ri t \epsilon(\theta) + \ri x p(\theta)$, lines with the right arrows are $f^{(-)}(\theta)\exp[-\ri t \epsilon(\theta) - \ri x p(\theta)]$. }
\label{cont}
\end{figure}

In \cite{akt} Eq. (\ref{kiev1}) was rewritten in the form which allowed an analytic continuation for real time. The $N$-th term in the square brackets of Eq. (\ref{kiev1}) is given as \cite{remark}  
\begin{widetext}  
\bea
\sum_{n=1}^{2N}\frac{1}{n!(2N-n)!}\int\prod_{i=1}^n\frac{\rd\theta_i}{2\pi} 
f^{(+)}(\theta_i)\re^{\tau\epsilon_i + 
\ri x p_i}\prod_{j=1}^{2N-n}\frac{\rd\theta'_j}{2\pi} f^{(-)}(\theta_j)\re^{ -\tau\epsilon_j - \ri x p_j}\frac{\prod_{i>k}\tanh^2[(\theta_i - \theta_k)/2]\prod_{j> p}\tanh^2[(\theta'_i - \theta'_p)/2]}{\prod_{i,p}
\tanh^2[(\theta_i - \theta'_p + \ri 0)/2]} \label{truth}
\eea
\end{widetext} 
where 
\bea
&& f^{(+)}(\theta_i) = \frac{\re^{\eta^{(+)}(\theta)}}{[\re^{\beta\epsilon(\theta)} -1]},  f^{(-)}(\theta_i) = \frac{\re^{-\eta^{(-)}(\theta)}}{[1 - \re^{-\beta\epsilon(\theta)}]}\label{eta}\\
&& \eta^{(\pm)}(\theta) = \frac{\ri M\sinh\theta}{\pi}\int_{-\infty}^{\infty}
\frac{\rd x \ln\{\coth[\beta\epsilon(x)/2]\}}{x^2 - M^2\sinh^2(\theta \pm \ri 0)} \nonumber
\eea
 and $p = M\sinh\theta, \epsilon = M\cosh\theta$. In Eq.(\ref{truth}), 
$n$ and $N-n$ are numbers of particles and antiparticles. Now we  rearrange the double sum (\ref{truth}) in such a way that we first sum all terms which contain a fixed difference between numbers of particles and antiparticles $2N -2n = 2k$. Then such term  in  Eq.(\ref{truth})  can  be represented as an integral of the correlation function of a Gaussian field theory:
\begin{widetext}
\bea
\frac{a_0^{-4k^2}}{(N-k)!(N+k)!}\int\prod_{i=1}^{N-k}\frac{\rd\theta_i}{2\pi a_0} 
f^{(+)}(\theta_i)\re^{\tau\epsilon_i + 
\ri x p_i}\prod_{j=1}^{N +k}\frac{\rd\theta'_j}{2\pi a_0} f^{(-)}(\theta_j)\re^{ -\tau\epsilon_j - \ri x p_j}\la \prod_i^{N - k}\re^{\ri\Phi(\theta_i + \ri a)}\prod_j^{N + k}\re^{-\ri\Phi(\theta_j' - \ri a)}\re^{2k\ri\Phi(\infty)}\ra_0 \label{corr}
\eea
\end{widetext}
where 
\bea
\la\Phi(\theta_1)\Phi(\theta_2)\ra \equiv G_0(\theta_{12}) = -\ln\left[\tanh^2(\theta_{12}) + a_0^2\right]
\eea
and $a >> a_0 \rightarrow 0$. Eq.(\ref{corr}) represents one term in the perturbative expansion of the partition function of the theory with the action 
\bea
&& S= \frac{1}{2}\int \rd\theta_1\rd\theta_2\Phi(\theta_1)G_0^{-1}(\theta_{12})\Phi(\theta_2) + \nonumber\\
&& \int \frac{\rd\theta}{2\pi a_0} V[\theta,\Phi(\theta)]\label{action}\\
&& V = f^{(+)}(\theta)\re^{ \tau M\cosh\theta +\ri x M\sinh\theta}\re^{\ri\Phi(\theta + \ri a)} + \nonumber\\
&& + f^{(-)}(\theta)\re^{ -\tau M\cosh\theta - \ri x M\sinh\theta}\re^{-\ri\Phi(\theta -\ri a)}
\eea
In this theory $x,t$ are external parameters. The field $\Phi$ lives on an infinite line in $\theta$ space.
It is easy to see that the entire correlation function can be written as 
\bea
&&\la\s(\tau,x)\s(0,0)\ra = \\
&& CM^{1/4}\re^{-|x|\Delta(T)}Z_0(x,t)
\mbox{lim}_{R \rightarrow \infty}\sum_{k=-\infty}^{\infty}\frac{\la\re^{2\ri k\Phi(R)}\ra_V}{a_0^{4k^2}}\nonumber\\
&& Z_0 = \int D\Phi \re^{-S[\Phi]}/\int D\Phi \re^{-S_0[\Phi]}\nonumber
\eea
and $\la...\ra_V$ stands for averaging with action (\ref{action}).  Already this form leads to significant simplifications. Namely, the partition function can be written as an exponent of the free energy $Z_0 = \exp[-F(x,t)]$. The latter one is represented as a sum of the cumulants:
\bea
&& F = -\sum_{N=1}^{\infty}\frac{1}{(N!)^2}\int\prod_{i=1}^{N}\frac{\rd\theta_i}{2\pi a_0} 
f^{(+)}(\theta_i)\re^{\tau\epsilon_i + 
\ri x p_i}\times\label{F}\\
&& \prod_{j=1}^{N}\frac{\rd\theta'_j}{2\pi a_0} f^{(-)}(\theta_j)\re^{ -\tau\epsilon_j - \ri x p_j}\la\la \prod_i^{N}\re^{\ri\Phi(\theta_i + \ri a)}\prod_j^N\re^{-\ri\Phi(\theta_j' - \ri a)}\ra\ra_0 \nonumber
\eea
where the subscript 0 stands for averaging with $V =0$ and
\begin{widetext} 
\bea
&&\la\re^{2\ri k\Phi(R)}\ra_V   = \sum_{N=|k|}^{\infty}\frac{1}{(N -k)!(N + k)!}\int\prod_{i=1}^{N+k}\frac{\rd\theta_i}{2\pi a_0} 
f^{(+)}(\theta_i)\re^{\tau\epsilon_i + 
\ri x p_i}\prod_{j=1}^{N -k}\frac{\rd\theta'_j}{2\pi a_0} f^{(-)}(\theta_j)\re^{ -\tau\epsilon_j - \ri x p_j}\times\nonumber\\
&&\la\la \prod_i^{N +k}\re^{\ri\Phi(\theta_i + \ri a)}\prod_j^N\re^{-\ri\Phi(\theta_j' -\ri a)}\re^{2\ri k\Phi(R)}\ra\ra_0 \label{corr1}
\eea
\end{widetext}

 All transformations so far have been exact. Now we would like to concentrate on the causal  Green's functions. To obtain them one has to replace $\tau$ with $\ri t$ in (\ref{truth}). Assuming that $T \ll M$, we consider the region of frequencies  $|\omega| < 2M$, where  the  only terms of the expansion (\ref{truth}) contributing to the spectral function are those which contain {\it equal} number of particles and antiparticles. As we shall demonstrate, all formfactor singularities are contained in the first term  of expansion (\ref{F}). Since $|f^{(+)}| \sim \exp(-\beta M) << 1, |f^{(-)}| \sim 1$, this expansion  is in powers of soliton density $\exp(-\beta M)$. The first term is given by

\begin{widetext}   
\begin{eqnarray}
&& -F^{(1)}(t,x) = \frac{1}{4\pi^2}\int \rd\theta_1\rd\theta_2 f^{(+)}(\theta_1)f^{(-)}(\theta_2)
\frac{\exp\left(\ri \{ t[\epsilon(\theta_1)- \epsilon(\theta_2)] + x[p(\theta_1)- p(\theta_2)]\}\right)}
{\tanh^2(\theta_{12} + \ri 0)/2} \nonumber\\
&& \approx \frac{1}{2\pi^2}\int \rd\theta\rd v f^{(+)}(\theta + v)f^{(-)}(\theta -v)
\frac{\exp\left[\ri  vM(t\sinh\theta +x\cosh\theta)\right]}
{(v + \ri 0)^2}\label{R}\\
&&\approx  \theta(t - |x|)\left\{\frac{1}{\pi}\int _{\tanh\theta < - |x|/t}\rd\theta f^{(+)}(\theta)f^{(-)}(\theta)M\left[(t\sinh\theta + x\cosh\theta)|\right] + \frac{2\ri}{\pi}g[\theta = \tanh^{-1}(x/t)]\right\}
\nonumber
\end{eqnarray}
\end{widetext}
where $
g(\theta) = f^{(+)}(\theta) - f^{(-)}(\theta)$ 
and, as follows from  
 (\ref{eta}) $
f^{(+)}(\theta)f^{(-)}(\theta) = \{4\cosh^2[\beta\epsilon(\theta)/2]\}^{-1}$.
 Here  we assumed that $Mt,(Tt)^{1/2}\gg [1-(x/t)^2]^{-1/2}$.
 The higher order cumulants contain higher powers of $\exp(-\beta M)$ and  also do not contain positive powers of $t$. This justifies keeping the exact distribution function in the real part of (\ref{R}). This cannot be done for the imaginary part, since  the second cumulant  gives a time independent contribution $\sim \exp(-2\beta M)$.  Within these limits  we obtain the following results:
\begin{widetext}
\bea
&&\la\s(x,t)\s(0,0)\ra_T  = 
 CM^{1/4}\theta(t)\exp(-\delta\Delta |x|)\exp\left\{- \frac{1}{4\pi}\int \rd p\frac{|tv(p) - x|}{\cosh^2[\beta\epsilon(p)/2]} - \frac{4\ri}{\pi}\exp[- \beta M/\sqrt{1 - (x/t)^2}]
\right\}, t > |x| \nonumber\\
&& \la\s(x,t)\s(0,0)\ra_T  = 
 CM^{1/4}\theta(t)\exp(-\Delta |x|), ~~ |x| > t \label{R1}\\
&&\la\mu(x,t)\mu(0,0)\ra_T  =\la\mu(x,t)\mu(0,0)\ra_{T=0}\la\s(x,t)\s(0,0)\ra_T \label{R2}\\
&& \delta\Delta = \frac{1}{\pi}\int \rd p\left[\ln\coth(\beta\epsilon/2) - \frac{1}{2\cosh^2(\beta\epsilon/2)}\right] \sim \exp(-3\beta M)\nonumber
\eea
\end{widetext}
 The imaginary part  of (\ref{R1})  in the time-like domain $t > |x|$ reflects a quantum nature of the excitations. For $T=0$ such imaginary part was first found in \cite{mccoy}. In the leading order in $\exp(-\beta M)$ Eq.(\ref{R2}) coincides with the one found in \cite{sachdev}.  At $x =0$ we have
\bea 
&& \la\s(x=0,t > 0)\s(0,0)\ra  \approx \label{asym}\\
&& CM^{1/4}\exp\left[- t/\tau_0 + \frac{4\ri}{\pi}\re^{-\beta M} + O\left(\re^{-2\beta M}\right)\right] \nonumber
\eea
where $\tau_0^{-1} = \frac{2T}{\pi}\frac{1}{\re^{\beta M} + 1}$. 
Since single solitons are not  directly observable, it is rather interesting to note that $\tau_0$ contains the distribution function of a single soliton. It would be very interesting to see what happens in correlation functions in models  containing particles with fractional statistics.

Since our approach shares certain common features  with the Fredholm determinant representation introduced by Korepin {\it et. al.} \cite{its},\cite{korepin}, \cite{comm},\cite{oota}, we feel obliged  to comment on the subject. The main difference is that  we do not represent the correlation functions as determinants though in certain limits this is possible. For instance, if one adopts  the nonrelativistic limit $\theta << 1$ in action (\ref{action}), it can be fermionized and rewritten as a theory of free fermions. Then by integrating  over fermions one  obtains the determinant representation. However, we would like to point out that such representation is not a goal in itself. By representing correlation functions as partition functions of some field theory one already achieves a lot since now one  can concentrate on connected diagrams where it is easier to keep track of singularities. It is possible that acting  along the lines of \cite{luk} one can obtain such representations  for interacting models. 

We also would like to warn against the direct comparison of Eq.(\ref{R1}) with  a similar equation for  the $\la \s^-\s^+\ra$ correlation function in the XY model obtained in \cite{its}. This warning  is necessary because the XY model in magnetic field is rather similar to the QI model; the similarity increases when the magnetic field exceeds the band width so that the ground state becomes ferromagnetic. However, as was explained to us by Korepin (private communication), the formulae of \cite{its} were obtained for the case of weak magnetic field the XY model spectrum is a gapless, which explains the difference in the final results.

This research was supported  by 
the DOE under contract number DE-AC02 -98 CH 10886.  We acknowledge valuable discussions with F. H. L. Essler, R. M. Konik, F. A. Smirnov, V. E. Korepin  and V. A. Fateev.


\begin{thebibliography}{99}
\bibitem{remark} In the paper \cite{akt} where this formula first appeared it was written in a simplified form valid in the limit $M >> T$.
\end{thebibliography}
\end{document}